\documentclass[twoside]{dis08}
\usepackage[latin1]{inputenc}
\usepackage[dvips]{graphicx,epsfig,color}
\usepackage{wrapfig,rotating}
\usepackage{amssymb,amsmath,array}

\begin{document}
\title{MC@NLO for Heavy Quarks in Photoproduction}

\author{Tobias Toll $^1$
%
%
\vspace{.3cm}\\
%
1- DESY \\
Notkestrasse 85, 22607 Hamburg - Germany
%
}


\maketitle
\newcommand{\dint}{\rm{d}}
\newcommand{\as}{\alpha_{\rm{s}}}

\begin{abstract}
An MC@NLO for heavy quarks in photoproduction is presented. This is the first lepton-hadron
process to be included into MC@NLO. To construct an MC@NLO process dependent so called 
MC-subtraction terms need to be calculated. The resulting calculation is compared
to a fixed order NLO calculation and the HERWIG event generator and is shown to perform
well. The slides can be found in \cite{url}.
\end{abstract}

\section*{Schematic Overview of the Calculation}
MC@NLO~\cite{mcatnlo1, mcatnlo2} is a generator for calculating
next-to-leading order matrix elements, combined  
with a Monte Carlo  event generator. When constructing an MC@NLO one needs to match the
 NLO matrix element 
with the parton shower of the generator. If wrongfully done, some of the configurations
produced by the real gluon emissions in the matrix element may also be produced 
by the first branching in the parton shower. Therefore, to avoid to doubly count 
these configurations the first branching in the parton shower is subtracted from the real 
emission part of the matrix element. This method is called 'modified subtraction', since 
it is based on the subtraction method of performing NLO calculations. The terms being 
subtracted are called 'Monte Carlo subtraction terms'. These terms are dependent upon which 
process is being calculated as well as on the parton shower itself. 
MC@NLO has 
previously been constructed for many processes, all in hadron-hadron collisions. In the 
following will be presented an MC@NLO for heavy quarks in photo production which is the 
first MC@NLO to be constructed for lepton-hadron collisions. The NLO calculation is taken
from FMNR~\cite{FMNR} and the parton shower from the HERWIG
event generator~\cite{HERWIG1} . 

\begin{figure}[b]
  \begin{center}
    \includegraphics[angle=0., scale=.5]{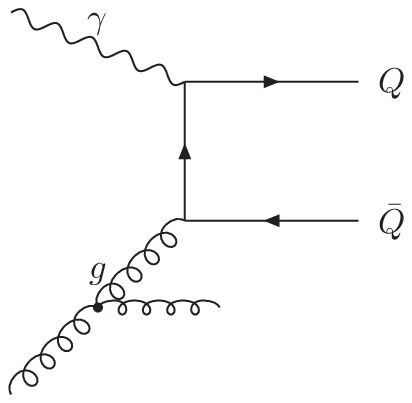}
    \includegraphics[angle=0., scale=.5]{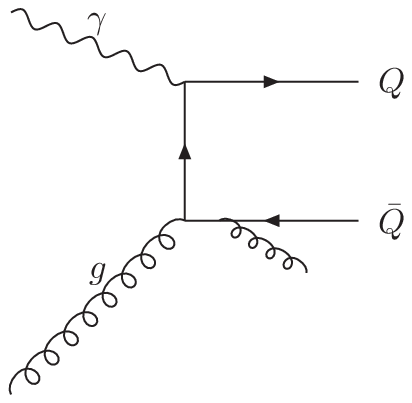}
    \includegraphics[angle=0., scale=.5]{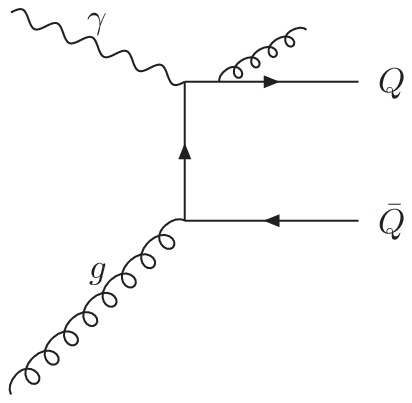}
    \includegraphics[angle=0., scale=.5]{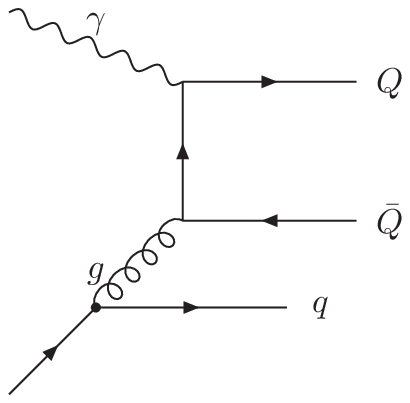}
    \includegraphics[angle=0., scale=.5]{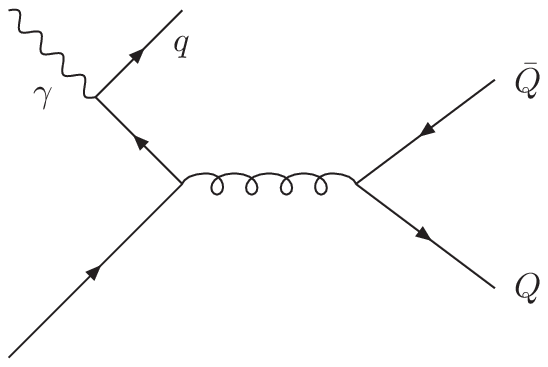}
  \end{center}
  \caption{Diagrams of the first possible branching in the HERWIG parton shower. 
    Charge conjugated diagrams are implied.}
  \label{fig:real1}
\end{figure}
For the NLO matrix element in this process, there are three kinds of contributions, 
namely the Born term, the real emissions, and the virtual corrections. 
The total NLO rate is schematically calculated as
\begin{eqnarray}
  \sigma_{\rm NLO}=\int \dint\phi_3 \left(\dint\sigma^{\rm Real}\right)
  +\int \dint\phi_2\left(\dint\sigma^{\rm Born}+\dint\sigma^{\rm Virtual}\right),
\end{eqnarray}
where the integral is over the phase-space $\phi_n$ where $n$ indicates how many 
particles there are in the final state.
These two integrals are separately divergent, even though their sum is finite. To solve
the integrals  a term precisely canceling the divergencies is subtracted from the first 
integral and added to the second integral:
\begin{eqnarray}
  \sigma_{\rm NLO}=\int\dint\phi_3 \left(\dint\sigma^{\rm Real}-\dint\sigma^{\rm subtr.}_3\right) 
  +\int\dint\phi_2 \left(\dint\sigma^{\rm Born}+\dint\sigma^{\rm Virtual}+
  \dint\sigma^{\rm subtr.}_2\right).
  \label{nlosubtr}
\end{eqnarray}
This procedure is called the 'subtraction method'. It should be noted that when integrated
over respective phase-space the subtraction terms cancel exactly.

In order to match such a calculation with the parton shower without doubly counting any 
configurations the subtraction method in eq.~\eqref{nlosubtr} is modified. 
Each branching in the parton shower contributes by an order of $\as$
This means that only the first branching in the parton shower could give 
rise to double counting, since all other branchings give a prediction at orders higher than
NLO.
The terms canceling the divergencies present in the integrals are then replaced by so called 
'Monte Carlo subtractio terms',
which also subtract the first branching in the parton shower from the real emissions, and 
the non-branching probability to the second integral:
\begin{eqnarray}
  \sigma_{\rm NLO}=\int\dint\phi_3 \left(\dint\sigma^{\rm Real}-\dint\sigma^{\rm MC}_3\right)
  +\int\dint\phi_2 \left(\dint\sigma^{\rm Born}+\dint\sigma^{\rm Virtual}+
  \dint\sigma^{\rm MC}_2\right).
  \label{mcsubtr}
\end{eqnarray}
\begin{figure}[t]
  \includegraphics[angle=0, scale=.3]{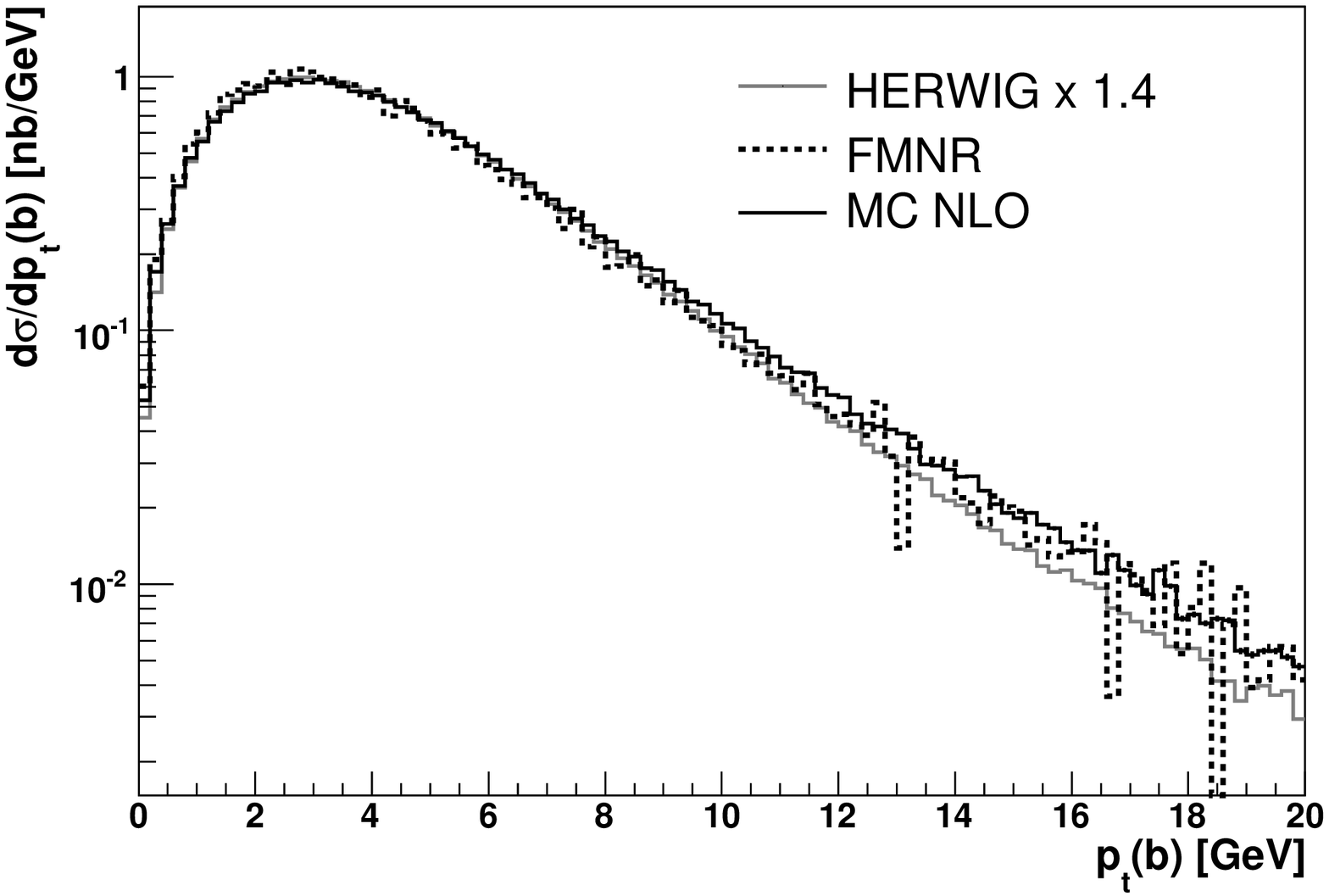}
  \includegraphics[angle=0, scale=.3]{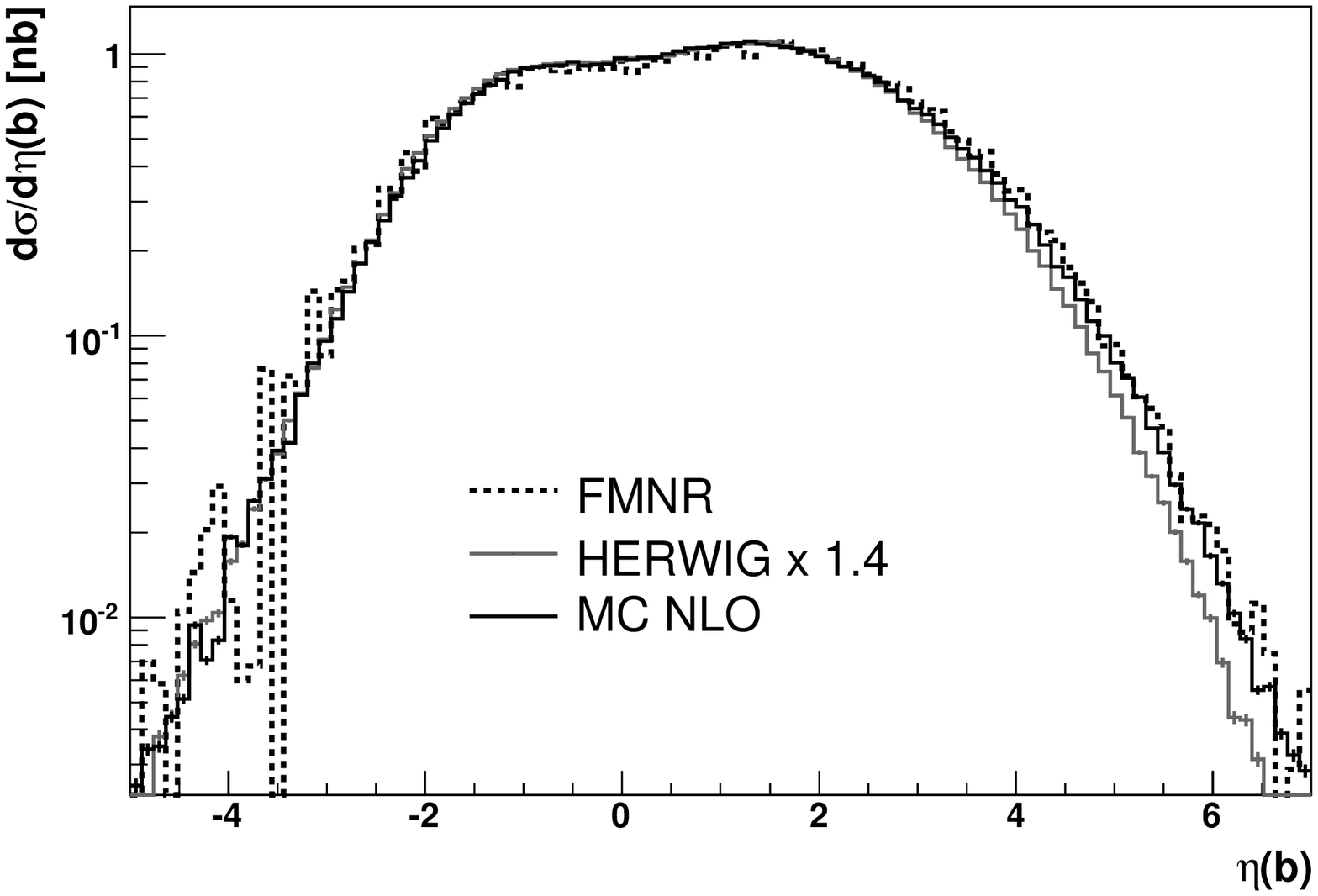}
  \caption{The $p_t$ and $\eta$ spectra of the b-quark as given by FMNR, 
    HERWIG and MC@NLO. Parameters are as in table \ref{tab:comp} and parton 
    densities are MRST2002NLO}
  \label{fig:pteta}
\end{figure}
The MC subtraction terms thus need to cancel all the intermediate divergencies of the 
calculation, and also ensure that the first branching in the parton shower is not counted 
twice, in such a way that the total NLO rate is still preserved. After integration, these
term do not cancel exactly, as was the case in eq.~\eqref{nlosubtr}, but this is 
compensated by the parton shower.

As explained above, to construct an MC@NLO, the MC-subtraction terms have to be calculated.
For heavy quarks in point-like photoproduction there are two types of processes to consider, 
namely:
\begin{eqnarray}
  \gamma g &\rightarrow& QQg \nonumber \\
  \gamma q/\bar{q} &\rightarrow& QQq/\bar{q}
\end{eqnarray}



The branchings neaded to be subtracted for these processes are shown in Fig.\ref{fig:real1}.
MC-subtraction terms for all these branchings have been calculated.

\begin{wraptable}{r}{0.7\columnwidth}
  \begin{center}
    \begin{tabular}{ | c | c | c | }
      \hline 
      Process & FMNR & MC@NLO \\ \hline
      Charm $[mb]$& $0.8423 \pm 0.003$ & $0.8423 \pm 0.003$ \\ \hline
      Beauty $[pb]$& $5215 \pm 10$ & $5199\pm 30$ \\\hline
    \end{tabular}
  \end{center}
  \caption{Comparisons of total prediction in FMNR and MC@NLO for charm and beauty 
    production. The errors are statistical. Here $m_c=1.5~GeV$, $m_b=4.95~GeV$ and  
    $\mu_{\rm R}=\mu_{\rm F} = m_{c/b, t}$. Parton densities are cteq5m}
  \label{tab:comp}
\end{wraptable}
The MC-subtraction terms need to cancel the divergencies present in the NLO-calculation. 
In order to test this, a regularized version of the  MC-subtraction terms are made
to approach the divergent soft and collinear limits, and are compared to a 
regularized version of the matrix element in the limit. This test has been performed 
succesfully.


\begin{wrapfigure}{r}{0.45\columnwidth}
  \begin{center}
    \includegraphics[angle=0, scale=.3]{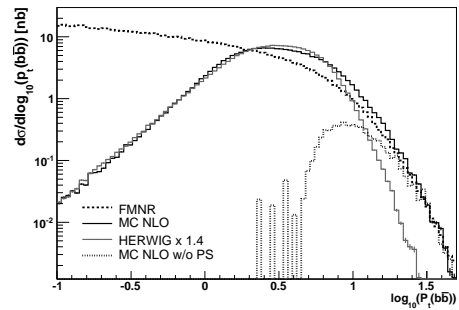}
  \end{center}
  \caption{The logarithm of the combined $p_t$ of the heavy quark pair as
    given by FMNR, HERWIG 
  and MC@NLO. Also shown is MC@NLO without parton showers. Parameters are as in 
  Fig.\ref{fig:pteta}.}
  \label{fig:logpt}
\end{wrapfigure}
To be sure that the double counting is properly canceled, by definition the total rate
given by MC@NLO must coinside with that from the NLO calculation. The predictions for 
charm and beauty production is compared in table \ref{tab:comp}.

\section*{Results}
The results from the new MC@NLO for heavy quarks in
$eP$-photoproduction are compared with predictions from HERWIG and FMNR. 
Only results on parton level are shown, and only from the point-like part 
of the calculation. The HERWIG predictions have been multiplied by a K-factor 
of $~1.4$. 

In Fig.\ref{fig:pteta} 
the $p_t$ and $\eta$ of the $b$-quark is shown. For these observables no large differences
between the calculations are expected, since they are not very dependent upon NLO effects.
However, as seen in Fig.\ref{fig:pteta} the MC@NLO prediction is always very close to that
of FMNR. Small differences are observed between HERWIG and the other calculations, 
especially in the very forward and backward parts of the $\eta$ distributions and 
for large $p_t>10~GeV$.

An observables where NLO effects are expected to be large are the combined $p_t$ of
the heavy quarks. By momentum conservation, this is equal to the combined  $p_t$ of all 
other final state partons. The results of this observable is shown in Fig.\ref{fig:logpt}.
In the case of FMNR, for small $p_t$ this observable show the $p_t$ distribution of the 
light parton emitted by the real emission, which diverges for small $p_t$. 
At $p_t(b\bar b)=0$ (not shown in the figure) the virtual terms instead dominate giving
a negative cross-section. It is therefore evident that the NLO-calculation is not sufficient
to describe this quantity. HERWIG, on the other hand, has resummation to all orders via the
parton shower. But the parton shower is still unable to describe the hardest part of
the $p_t$-tail for the heavy quark pair. MC@NLO also has resummation to all orders via
the parton shower, but the NLO-matrix element also manages to reproduce the hard tail.
Shown in Fig.\ref{fig:logpt} is also the prediction of MC@NLO {\it without} parton showers. 
This shows how the MC-subtraction is different from the NLO subtraction in FMNR. 
It should be noted that the total cross-section in all four calculations are equal.
\begin{wrapfigure}{r}{0.45\columnwidth}
  \begin{center}
    \includegraphics[angle=0, scale=.3]{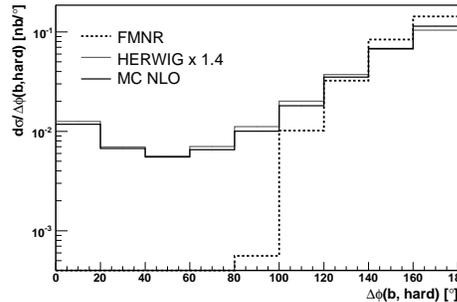}
  \end{center}
  \caption{The difference in azimuthal angle $\phi$ between the $b$-quark and the 
    hardest other parton.}
  \label{fig:deltaphi}
\end{wrapfigure}
Another observable where NLO effects are expected to be large is the difference in
azimuthal angle $\phi$ between one of the heavy quarks and the hardest other parton. 
The result for this observable is shown in Fig. \ref{fig:deltaphi}. At LO, when no
light parton is emitted, the heavy quarks are always back-to-back. In FMNR,
for the light parton to be harder than the heavy anti-quark the angle needs to be 
larger than $90^\circ$. In HERWIG there are almost always more than three partons available
and thus may the hardest other parton have any angle relative to the heavy quark. 
As seen in the figure, this is also the case for MC@NLO.

\section*{Conclusions}
The first version of MC@NLO for heavy quarks in photoproduction has been presented. 
The MC-subtraction terms have been constructed 
and successfully tested. Also, the predictions
at parton level agree with that of FMNR whenever NLO is expected to be important, and 
with that of HERWIG whenever 
the parton shower is expected to be important.
The full MC@NLO for heavy quarks in photoproduction is expected to be presented soon, 
including parton showers and hadronization. This will be a tool for comparison with data
from the HERA experiments.

\section*{Acknowledgments}
I would especially like to thank Stefano Frixione and Bryan Webber who both have 
contributed a great deal to the construction of the MC@NLO program here described.
Stefano has also contributed with invaluable advice during the process as has Hannes Jung
to whom I am also very grateful. I would like to thank the organizers of DIS08 for a very 
interesting and inspirational conference. 
I am especially grateful for the invitation to contribute with this note to the DIS proceeding. 


\begin{footnotesize}


\bibliographystyle{unsrt}
\bibliography{toll_tobias.bib}
%

\end{footnotesize}

\end{document}